\documentclass[conference,compsoc]{IEEEtran}
\usepackage{cite}
\usepackage{spconf}

  \usepackage{graphicx}
\makeatletter

\newtheorem{theorem}{Theorem}
\newtheorem{corollary}{Corollary}
\newcommand{\Rmnum}[1]{\expandafter\@slowromancap\romannumeral #1@}
\makeatother

\usepackage{amsmath}
\usepackage{bm}
\usepackage{amssymb}
\ifCLASSOPTIONcompsoc
  \usepackage[caption=false,font=footnotesize,labelfont=bf,textfont=normalfont]{subfig}
\else
  \usepackage[caption=false,font=footnotesize]{subfig}
\fi
\DeclareMathOperator{\tr}{Tr}

\begin{document}
\title{DESIGN OF SAMPLING SET FOR BANDLIMITED GRAPH SIGNAL ESTIMATION}

%\author{\IEEEauthorblockN{Xuan Xie, Hui Feng, Bo Hu}
%\IEEEauthorblockA{Department of Electronic Engineering,
%Fudan University, Shanghai 200433, China\\
%Emails: \{xxie15, hfeng, bohu\}@fudan.edu.cn}}
\name{Xuan Xie, Hui Feng, Junlian Jia, Bo Hu}
\address{Research Center of Smart Networks and Systems,
Fudan University, Shanghai 200433, China\\
%Research Center of Smart Networks and Systems\\
Emails: \{xxie15, hfeng, jljia10, bohu\}@fudan.edu.cn}
%\and
%\IEEEauthorblockN{Homer Simpson}
%\IEEEauthorblockA{Twentieth Century Fox\\
%\and
%\IEEEauthorblockN{James Kirk\\ and Montgomery Scott}
%\IEEEauthorblockA{Starfleet Academy\\
%San Francisco, California 96678--2391\\


%Telephone: (800) 555--1212, Fax: (888) 555--1212}


\maketitle

\begin{abstract}

It is of particular interest to reconstruct or estimate bandlimited graph signals, which are smoothly varying signals defined over graphs, from partial noisy  measurements. However, choosing an optimal subset of nodes to sample is NP-hard. We formularize the problem as the experimental design of a linear regression model if we allow multiple measurements on a single node. By relaxing it to a convex optimization problem, we get the proportion of sample for each node given the budget of total sample size. Then, we use a probabilistic quantization to get the number of each node to be sampled. Moreover, we analyze how the sample size influences whether our object function is well-defined by perturbation analysis. Finally, we demonstrate the performance of the proposed approach through various numerical experiments.
\end{abstract}

% \fi
\IEEEpeerreviewmaketitle

\section{Introduction}
Graph signals (GS) are rather useful for describing signals and information in irregular domains, such as social, sensor and economic networks \cite{shuman2013emerging}. Graph signal processing (GSP) theory extends and generalizes the classic discrete signal processing theory to graphs by introducing graph Fourier transform \cite{chen2016signal,anis2016efficient,sandryhaila2013discrete,sandryhaila2014discrete,
narang2013signal,chen2015discrete,marques2016sampling,puy2016random,
tsitsvero2016signals,dong2014clustering,wang2015local,shi2015infinite,
girault2015stationary,perraudin2017stationary}, wavelet \cite{hammond2011wavelets,narang2013compact,shuman2015spectrum,tremblay2014graph,narang2012perfect}, etc. Two main approaches have been considered to construct a graph Fourier analysis framework: algebraic graph theory linked to the adjenacy matrix \cite{chen2016signal,sandryhaila2013discrete,sandryhaila2014discrete,narang2013signal,chen2015discrete,marques2016sampling} and spectral graph theory based on the graph Laplacian matrix \cite{anis2016efficient,
dong2014clustering,wang2015local,marques2016sampling,shi2015infinite,girault2015stationary,
perraudin2017stationary,puy2016random,tsitsvero2016signals}. GSP theory have been developed to graph filter \cite{narang2012perfect,narang2013compact,shi2015infinite}, community detection and clustering on graphs \cite{tremblay2014graph,dong2014clustering}, sampling and interpolation \cite{chen2015discrete,narang2013signal,anis2016efficient,puy2016random,tsitsvero2016signals} and corresponding distributed algorithm \cite{wang2015local,marques2016sampling}. Recently, stationary processes to the graph domain has been investigated in \cite{girault2015stationary,perraudin2017stationary}.

Sampling theory for GS deals with the problem of recovering a bandlimited signal from its samples on a subset of nodes of the graph. Bandlimited GS widely exist in most real-world networks due to the fact that the signals on adjacent nodes won't differ dramatically in reality. To formulate a sampling theory for bandlimited GS, the following questions arises: How to choose the best sampling set which can achieve the smallest reconstruction error? Generally, the signals can be bandlimited or approximate bandlimited and the samples can be noise-free or noisy as shown in Table. \ref{literature}. In this paper, we focus on choosing the sampling set for bandlimited GS from noisy measurements. 

\begin{table}[!h]
\renewcommand{\arraystretch}{1.5}
% if using array.sty, it might be a good idea to tweak the value of
% \extrarowheight as needed to properly center the text within the cells
\caption{Literature based on different assumptions}
\label{literature}
\centering
% Some packages, such as MDW tools, offer better commands for making tables
% than the plain LaTeX2e tabular which is used here.
\begin{tabular}{l c c}
\hline
 & Noise-free & Noisy\\
\hline
Bandlimited & \cite{chen2015discrete,narang2013signal} & \cite{chen2015discrete,anis2016efficient,wang2015local,marques2016sampling,puy2016random,tsitsvero2016signals}\\
\hline
Approximate bandlimited & \cite{narang2013signal} & \cite{anis2016efficient,chen2016signal}\\
\hline
\end{tabular}
\end{table}

In this work, we connect the design of sampling set for noisy bandlimited GS with experimental design problem, which is closely related to the optimal experimental design problem in \cite{fedorov1972theory,pukelsheim2006optimal}. This connection with experimental design has also been noticed in \cite{anis2016efficient, marques2016sampling} for GS sampling but not fully investigated. The original form of experimental design problem with given measurement size is a combinatorial problem and NP-hard. There are two main kinds of solutions. One is heuristic algorithms, which is time-consuming and can hardly produce an optimal solution, such as Fedorov’s exchange algorithm \cite{miller1994algorithm} and Wynn’s algorithm \cite{wynn1972results}. The other is the convex relaxation proposed in \cite[Sec. 7.5]{boyd2004convex}， which tries to obtain an optimal solution. It has been also claimed to give excellent results in many applications \cite{joshi2009sensor}. In this paper, we follow the latter.

In this work, we consider the case when any node is allowed to be sampled multiple times. From the view of experimental design, the optimal solution usually achieves when we allow multiple-time sampling. This is very feasible for many real-world networks such as sensor network and social network. We not only decide the nodes to be sampled but also get the proportion of sample for each node of a given sample size by relaxing the combinational problem to convex optimization. In order to get the number of each node to be sampled, we use probabilistic quantization \cite{aysal2007distributed} to unbiased quantify the solution of the relaxed problem to integers to get a suboptimal solution of the combinational problem. Different from the heuristic algorithm \cite{anis2016efficient, marques2016sampling} whose performance can not be evaluated, we take the quantization error as the perturbation and analyze how sample size influence the performance of our algorithm by perturbation analysis. Moreover, we find a lower bound of the sample size to ensure the object function of our algorithm well-defined, which can provide a reference to practical problems. Finally, the performance of our method is analyzed and shown to have smaller reconstruction error and is more robust against additive noise. 

\section{System Model}
 Consider an $N$-vertex undirected connected graph $\mathcal{G}=(\mathcal{V},\mathcal{E},\bm{W})$, where $\mathcal{V}$ is the vertex set, $\mathcal{E}$ is the edge set, and $\bm{W}$ is the weighted adjacency matrix. If there is an edge $e=(i,j)$ between vertices $i$ and $j$, then the entry $W_{i,j}$ represents the weight of the edge; otherwise $W_{i,j}=0$. A signal $\bm{f}$ defined on the vertices of a graph can be represented as a vector $\bm{f} \in \mathbb{R}^N$, and its element $f_i$ represents the signal value at the $i$th vertex in $\mathcal{V}$.

 The graph Laplacian is defined as $\bm{L}=\bm{D}-\bm{W}$, where the degree matrix $\bm{D}=\text{diag}(\bm{1W})$. Since the Laplacian matrix is real symmetric, it has a complete eigenbasis and the spectral decomposition 
\begin{IEEEeqnarray}{Rl}
\label{spectral decomposition}
\bm{L}=\bm{V}\bm{\Lambda}\bm{V}^T,
\end{IEEEeqnarray}
where the eigenvectors  $\{v_k\}_{0\leq k\leq N-1}$ of $\bm{L}$ form the columns of $\bm{V}$, and $\bm{\Lambda}
\in \mathbb{C}^{N\times N}$ is a diagonal matrix of eigenvalues $0=\lambda_0\leq \lambda_1\leq \cdots\leq \lambda_{N-1}$ of $\bm{L}$. The graph Fourier transform corresponds to the basis expansion of a signal. The eigenvectors of the graph Laplacian are regarded as the Fourier basis and the eigenvalues are regarded as frequencies \cite{sandryhaila2014discrete}. The expansion coefficients of a graph signal $\bm{f}$ in terms of eigenvectors are defined as $\hat{\bm{f}}$, so that a graph signal can be expressed as
\begin{IEEEeqnarray}{Rl}
\label{GFT}
\bm{f}=\bm{V}\hat{\bm{f}}.
\end{IEEEeqnarray}
A graph signal is called \emph{bandlimited} when there exists a $K\in \{0,1,\cdots,{N-1}\}$ such that its graph Fourier transform $\hat{\bm{f}}$ satisfies $\hat{f}_k=0$ for all $k\geq K$\cite{chen2015discrete}. Let $\bm{f}$ be a signal with bandwidth $K$, then it satisfies
\begin{equation}
\label{bandlimited signal}
\bm{f}=\bm{V}_{K}\hat{\bm{f}}_{K},
\end{equation}
where $\bm{V}_{K}$ denotes the first $K$ columns of $\bm{V}$ and $\hat{\bm{f}}_{K}$ denotes the first $K$ coefficients of $\hat{\bm{f}}$.

Suppose that we sample $M$ measurements from the graph signal $\bm{f}\in \mathbb{C}^N$ to produce a sampled signal $\bm{f}_{\mathcal{S}}\in \mathbb{C}^M$, usually $M\leq N$, where $\mathcal{S}=(\mathcal{S}_1,\cdots,\mathcal{S}_M)$ denotes the sequence of sampled indices, and $\mathcal{S}_i\in \{1,2,\cdots,N \}$. The sampling operator $\Psi:\mathbb{C}^N\mapsto \mathbb{C}^M$ is defined as
\begin{equation}
\label{sampling operator}
\Psi_{i,j}=
        \left\{\begin{matrix}
        1, & j=\mathcal{S}_i;\\
        0, & \text{otherwise.}
        \end{matrix}\right.
\end{equation}
Let $\bm{w}\in \mathbb{C}^M$ be the $i.i.d.$ noise with zero mean and unit variance introduced during sampling. Then the samples are given by $\bm{f}_{\mathcal{S}}=\Psi \bm{f}$, and the observation model is
$\bm{y}_{\mathcal{S}}=\Psi \bm{f}+\bm{w}$.
Considering (\ref{bandlimited signal}), for bandlimited GS, the observation model can be expressed as $\bm{y}_{\mathcal{S}}=\Psi\bm{V}_{K}\hat{\bm{f}}_{K}+\bm{w}$. Let $\bm{V}_{MK}=\Psi\bm{V}_{K}$, then 
\begin{IEEEeqnarray}{Rl}
\bm{y}_{\mathcal{S}}=\bm{V}_{MK}\hat{\bm{f}}_{K}+\bm{w}. 
\end{IEEEeqnarray}
The best linear unbiased estimation \cite{kay1993fundamentals} of $\hat{\bm{f}}_{K}$ from observed samples $\bm{y}_{\mathcal{S}}$ is,
\begin{equation}
\bm{\hat{f}}'_K=\bm{V}_{MK}^{\dagger}\bm{y}_\mathcal{S},
\end{equation}
where $\bm{V}_{MK}^{\dagger}=(\bm{V}_{MK}^T\bm{V}_{MK})^{-1}\bm{V}_{MK}^T$ is the pseudo-inverse of $\bm{V}_{MK}$.

By (\ref{GFT}), the estimation error of $\bm{f}$ is
\begin{IEEEeqnarray}{Rl}
\label{estimation error}
\bm{e}=\bm{f}'-\bm{f}=\bm{V}_{K}\bm{V}_{MK}^{\dagger}\bm{w}.
\end{IEEEeqnarray}
The covariance matrix of estimation error is
\begin{IEEEeqnarray}{Rl}
\label{covariance matrix}
\bm{E}=\mathbb{E}[\bm{e}\bm{e}^T]=\bm{V}_{K}\left(\bm{V}_{MK}^\mathrm{T}\bm{V}_{MK}\right)^{-1}\bm{V}_{K}^T.
\end{IEEEeqnarray}
Our main problem is to choice an optimal $\Psi$ that minimize the error covariance $\bm{E}$ in certain scalarization forms. 

Several scalarizations of (\ref{covariance matrix}) have been proposed for the convenience of minimization as follow:\\
%\begin{itemize}
%\item 
$D$-\textbf{optimal}
\begin{equation}
f(\bm{E}) = \log \det\left(\bm{E}\right),
\end{equation}
%\item 
$E$-\textbf{optimal}
\begin{equation}
f(\bm{E}) = \left\|\bm{E}\right\|_2,
\end{equation}
%\item 
$A$-\textbf{optimal}
\begin{equation}
f(\bm{E}) =  \tr\left(\bm{E}\right).
\end{equation}
%\end{itemize}
It is obvious that $f(\bm{E})=f\left(\left(\bm{V}_{MK}^\mathrm{T}\bm{V}_{MK}\right)^{-1}\right)$ for all the three scalarizations above, since $\bm{V}_{K}$ is  orthogonal.

\section{Algorithm}
In this section, we consider the sampling set design problem by estimating $\bm{f}$ from measurements. Let $\bm{u}_1^T,\cdots,\bm{u}_N^T$ be the rows of $\bm{V}_{K}$. Recall from $\bm{V}_{MK}=\Psi\bm{V}_{K}$ that the the rows of $\bm{V}_{MK}$, which characterize the measurements, can be chosen among $N$ possible test vectors $\bm{u}_1^T,\cdots,\bm{u}_N^T$. Our goal of optimal experimental design is to make $f(\bm{E})$ as small as possible.

Let $m_i$ denotes the number of experiments for which $\bm{u}_i$ is chosen, and assume that the sample size is $M$, so we have
\begin{equation}
m_1+\cdots+m_N=M.
\end{equation}
The scalarization of error covariance matrix can be expressed as
\begin{IEEEeqnarray}{Rl}
f(\bm{E})%= & \left(\bm{V}_{MK}^\mathrm{T}\bm{V}_{MK}\right)^{-1}
	%= \left(\sum_{i=1}^{M}\bm{a}_i \bm{a}_i^T\right)^{-1}\IEEEnonumber\\
	= f\left(\left(\sum_{i=1}^{N}m_i\bm{u}_i \bm{u}_i^T\right)^{-1}\right).
\end{IEEEeqnarray}
It shows that the error covariance depends only on the numbers of each $\bm{u}_i$ is chosen.
The basic experimental design problem is as follow,
\begin{IEEEeqnarray}{Rl}
\label{basic experiment}
\underset {m_i} {\text{minimize}} \quad \quad  & f(\bm{E})\IEEEnonumber\\%=\left(\sum_{j=1}^{N}m_j\bm{u}_j \bm{u}_j^T\right)^{-1}
         \text{subject to} \quad \quad  & m_i\geq 0, \quad m_1+\cdots+m_N=M\IEEEnonumber\\
        & m_i\in \bm{Z}.
\end{IEEEeqnarray}

\subsection{THE RELAXED PROBLEM}
The basic experimental design problem (\ref{basic experiment}) is an intractable combinatorial problem. We relax the constraint that the $m_i$ are integers following Boyd's method \cite[Sec. 7.5]{boyd2004convex}. Let $p_i=m_i/M$, which indicates the proportion of experiment $i$, and relax the constraint that each $p_i$ is an integer multiple of $\frac{1}{M}$, we obtain the relaxed experimental design problem
\begin{IEEEeqnarray}{Rl}
\label{relaxed problem}
\underset {p_i} {\text{minimize}} \quad \quad  & f(\bm{E}) = f\left(\left(\sum_{i=1}^{N}p_i\bm{u}_i \bm{u}_i^T\right)^{-1}\right)\IEEEnonumber\\
         \text{subject to} \quad \quad  & \bm{p}\succeq 0,\quad \bm{1}^T\bm{p}=1.
\end{IEEEeqnarray}
This is a convex optimization which can be solved by any optimization tool like interior-point methods \cite{boyd2004convex}. In the rest of this paper, we consider only the relaxed experimental problem.
The optimal value of the relaxed problem (\ref{relaxed problem}) provides a lower bound on the optimal value of the combinatorial one since the combinatorial problem has an addition constraint.

After solving (\ref{relaxed problem}), each entry of $\bm{p}$ need to be quantified to an integer multiple of $\frac{1}{M}$. Different form Boyd's method \cite[Sec. 7.5]{boyd2004convex}, we use probabilistic quantization \cite{aysal2007distributed} instead of the simple rounding to ensure an unbiased quantization in mean. The probabilistic quantization $Q:p_i\to Q(p_i)$ is defined as follow: $p_i\in [0,1]$ is equally divide into $M-1$ sub-intervals. The quantization points is defined as $\{1/M,2/M,\ldots,1\}$. Then, for $p_i\in [k/M,(k+1)/M]$, $k\in\{0,1,\ldots,M-1\}$, $Q(p_i)$ is a random variable defined by
\begin{IEEEeqnarray}{Rl}
\label{quantization solution}
Q(p_i)=\begin{cases}
 \frac{k}{M}& \text{ with probability } (\frac{k+1}{M}-p_i)M \\
\frac{k+1}{M} & \text{ with probability } (p_i-\frac{k}{M})M.
\end{cases}
\end{IEEEeqnarray}

Clearly we have $\left|p_i-Q(p_i)\right|\leq1/(2M)$. So when $M$ is large enough, we have $\bm{p}\approx Q(\bm{p})$, which implies the error covariance matrix associated with $\bm{p}$ and $Q(\bm{p})$ are closed.

As a result, we can use $Q(\bm{p})$ to generate a suboptimal sampling set: let $m_i = MQ(\bm{p}_i)$ be the sample quota of the $i$th node.

\subsection{PERTURBATION ANALYSIS OF QUANTIZATION}

By relaxing the experimental design problem, we do not need to solve the combinatorial problem directly. Meanwhile, the suboptimal solution brings a new problem of how to ensure the objective function of (\ref{relaxed problem}) invertible when we using $Q(p_i)$ to replace $p_i$, since the object function will be ill-condition if $\sum_{i=1}^{N}Q(p_i)\bm{u}_i \bm{u}_i^T$ is not invertible.

Let $\bm{A}=\sum_{i=1}^{N}p_i\bm{u}_i \bm{u}_i^T$ and $\bm{\hat{A}}=\sum_{i=1}^{N}Q(p_i)\bm{u}_i \bm{u}_i^T$, our goal is to ensure $\hat{\bm{A}}$, perturbed from $\bm{A}$, invertible. Let the perturbation error $\Delta p_i=Q(p_i)-p_i$, and the perturbation is given by
\begin{IEEEeqnarray}{Rl}
	\label{deltaA}
	\delta \bm{A} = \hat{\bm{A}}-\bm{A}=\sum_{i=1}^{N}\Delta p_i\bm{u}_i \bm{u}_i^T.
\end{IEEEeqnarray}

According to the \cite[Th 2.1]{demmel1997applied}, known that $\bm{A}$ is nonsingular, the relative $l_2$-norm distance from $\bm{A}$ to the nearest singular matrix is
\begin{equation}
\min \left\{ \frac {\left\|\left(\delta \bm{A}\right)\right\|_2} {\left\|\bm{A}\right\|_2} \text{:} \bm{A}+\delta \bm{A}\; \text{singular}\right\} =\frac {1} {\left\|\bm{A}\right\|_2\left\|\bm{A}^{-1}\right\|_2}.
\end{equation}
Therefore, to ensure $\hat{\bm{A}}$ invertible, the following condition needs to be satisfied,
\begin{IEEEeqnarray}{Rl}
\label{condition}
 \left\|\delta \bm{A}\right\|_2 & < \frac {1} {\left\|\bm{A}^{-1}\right\|_2}=\sigma_{\min}(\bm{A}).
\end{IEEEeqnarray}
Recall from (\ref{deltaA}) that $\left\|\delta \bm{A}\right\|_2$ is related to $\Delta p_i$. Thus, $\Delta p_i$ affects the probability of (\ref{condition}) being held. 
\begin{theorem}
Suppose that $\Delta p_1,\ldots,\Delta p_N$ are independent, and each $Q(p_i)$ is obtained by the quantization given in (\ref{quantization solution}). The probability of (\ref{condition}) is 
\begin{equation}
P(\left\|\delta \bm{A}\right\|_2 < \sigma_{\min}(\bm{A})) > \prod_{i=1}^N\left(1-\frac{\text{Var}[\Delta p_i]}{(\sigma_{\min}(\bm{A}))^2}\right).
\label{probability}
\end{equation}
\end{theorem}

\begin{IEEEproof}
Let $\Delta\bm{P} = \text{diag}(\Delta p_1,\Delta p_2,\ldots,\Delta p_N)$, then
\begin{IEEEeqnarray}{Rl}
 \label{conditionnum}
	\left\|\delta\bm{A}\right\|_2 = & \left\| \bm{V}_k^T \Delta \bm{P}\bm{V}_k \right\|_2\IEEEnonumber\\
	 \leq & \left\|\bm{V}_k^T \right\|_2\left\|\Delta \bm{P}\right\|_2 \left\|\bm{V}_k \right\|_2
	 =  \max  \left|\Delta p_i\right|.
\end{IEEEeqnarray}
Since $\Delta p_1,\ldots,\Delta p_N$ are independent, we can get the following inequation,
\begin{IEEEeqnarray}{Rl}
P(\left\|\delta \bm{A}\right\|_2 < \sigma_{\min}(\bm{A})) \geq \prod_{i=1}^N P(\left|\Delta p_i\right| < \sigma_{\min}(\bm{A})).
\end{IEEEeqnarray}
According to Chebycheff inequality, for any $p_i\in [k/M,(k+1)/M]$, the following inequation holds,
\begin{IEEEeqnarray}{Rl}
P(\left|\Delta p_i\right| < \sigma_{\min}(\bm{A}))>1-\frac{\text{Var}[\Delta p_i]}{(\sigma_{\min}(\bm{A}))^2}.
\end{IEEEeqnarray}
Thus proving our claim.
\end{IEEEproof}

Assume that $M$ is large enough, then any $p_i\in [k/M,(k+1)/M]$ is approximated to be a uniform distribution: $p_i\sim U[k/M,(k+1)/M]$. %Denote $\Delta_1=\left(\frac{k}{M},\frac{2k+1}{2M}\right]$ and $\Delta_2=\left(\frac{2k+1}{2M},\frac{k+1}{M}\right]$.%
According to (\ref{quantization solution}), for every $p_i\in [k/M,(k+1)/M)$, $\mathbb{E}[Q(p_i)] = p_i$, which means $Q(p_i)$ is an unbiased representation of $p_i$, so
\begin{IEEEeqnarray}{Rl}
\text{Var}[\Delta p_i]
%= \mathbb{E}[{\Delta p_i}^2]
= \int_{-\frac{1}{2M}}^{\frac{1}{2M}}{\Delta p_i}^2 \text{p}(\Delta p_i)\text{d}\Delta p_i
\approx &\frac{5}{192M^3}.
\end{IEEEeqnarray}

We do not need the probability in (\ref{probability}) to be 1 most of the time, it can be reduced to $\eta(0<\eta<1)$ according to actual needs.
This leads to the following corollary.
\begin{corollary}
$\bm{\hat{A}}=\sum_{i=1}^{N}Q(p_i)\bm{u}_i \bm{u}_i^T$ is invertible with probability $\eta$ if the sample size $M$ satisfies:
\begin{equation}
\label{sample size}
M \geq \left \lceil \left(\frac{5}{192(1 - \sqrt[N]{\eta})(\sigma_{\min}(\bm{A}))^2}\right)^{\frac{1}{3}} \right \rceil,
\end{equation}
where $\lceil \cdot \rceil$ denotes the ceiling operation.
\end{corollary}

\section{Simulation}
%\begin{figure}[!t]
%  \centering
%  \includegraphics[width=5cm]{SW_error_cut}
%\end{figure}
%\begin{figure}[!t]
%  \centering
%  \includegraphics[width=3.7cm]{SW_snr_cut}
%\caption{Reconstruction results for different graphs and signals. (a) Graph G2 and Signal F1. (b) Graph G2 and Signal F2.}
%\label{SW_snr}
%\end{figure}

\begin{figure}[!t]
  \centering
	\subfloat[Signal F1]{\includegraphics[width=0.44\textwidth]{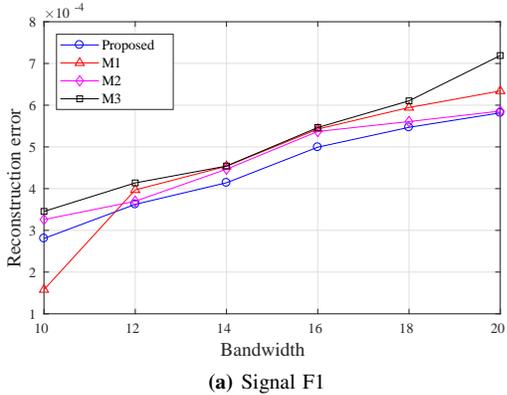}
	\label{error_SW}}\\
  \centering
	\subfloat[Signal F2]{\includegraphics[width=0.44\textwidth]{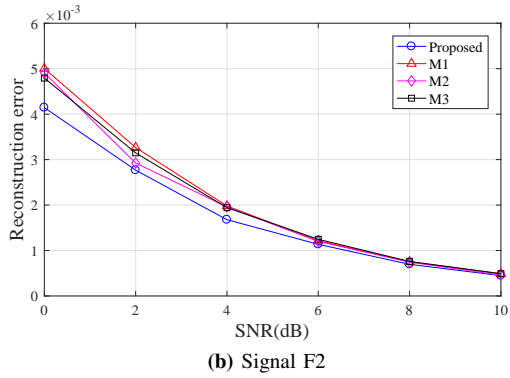}
	\label{snr_SW}}
\caption{Reconstruction results of G1 for different signals.}
\label{SW_snr}
\end{figure}

We now numerically evaluate the performance of the proposed work. The experiments compares the average reconstruction error $\left\|\bm{e}\right\|_2$ of different sample set selection algorithms versus different bandwidth of the true signal and different SNR of the samples. We compare our approach with the following methods: M1 \cite{chen2015discrete} and M2 \cite{anis2016efficient}, which are both greedy algorithm, as well as M3 \cite{joshi2009sensor}, which uses relaxed convex optimization to calculate the probability to sample each node and select $M$ nodes with largest probability.
%\begin{itemize}
%\item[M1]: This method \cite{chen2015discrete} uses a greedy method to approximate the $\mathcal{M}$ that maximizes $\sigma_{\min}(\bm{{V}_{MK}})$.
%\item[M2]: This method \cite{anis2016efficient} uses a greedy heuristic algorithm. At each step, this method computes the smoothest signal $\phi_k^*\in L_2(\mathcal{M})$(the value of $k$ is set to 8 in our simulation) and include the node on which the smoothest signal has maximum energy.
%\item[M3]: This method \cite{boyd2004convex} uses relaxed convex optimization to calculate the probability of being sampled of each node to achieve a minimum reconstruction error. And the nodes with $M$ largest probability was selected.
%\end{itemize}

We give some simulation results on the following simulated undirected graphs:
%\begin{itemize}
%\item[G1:]
Small world graph (G1) \cite{watts1998collective} of 1000 nodes rewiring with probability 0.1. 
%\item[G2:]
Random geometric graph (G2) \cite{perraudin2014gspbox} with 500 nodes placed randomly in the unit square and edges are placed between any nodes within 0.6. The edge weights are assigned via a Gaussian kernel.
%\end{itemize}

For each of the above graphs, we consider the problem in the following scenarios:
%\begin{itemize}
%\item[F1:]
(F1) The bandwidth of the true signal varying from 10 to 20 and non-zero GFT coefficients are generated from $\mathcal{N}(1,0.5^2)$. The samples are noisy with additive $i.i.d.$ Gaussian noise such that the SNR equals 10 dB. 
%\item[F2:]
(F2) The true signal is exactly with the bandwidth of 15 and non-zero GFT coefficients are generated from $\mathcal{N}(1,0.5^2)$. The samples are noisy with additive $i.i.d.$ Gaussian noise and the SNR varies from 0dB to 10dB.
%\end{itemize} 

%Firstly, we compare the performance of all methods when the bandwidth of the graph signal varies. We generate 200 signals from the signal model above on the each of the graphs. For the signal with a given bandwidth, we obtain the least sample size $M$ using (\ref{sample size}) and use the sample sets obtained from all the methods to perform reconstruction and plot the mean of reconstruction error. The result of reconstruction error is illustrated in Fig. \ref{compare_all_SW_K}. The result of and the sample size of all methods for each bandwidth and the corresponding number of nodes sampled of our method is illustrated in Fig. \ref{compare_all_SW_size}.
%
%Then, given the sample size $M=200$, we compare the performance of all methods when the SNR of samples varies.  The result of reconstruction error is illustrated in Fig. \ref{compare_all_SW_snr}. The result of the sample size and the corresponding number of nodes sampled of our method for each bandwidth is illustrated in Fig. \ref{compare_all_SW_snr_size}.
\begin{figure}[!t]
  \centering
	\subfloat[Signal F1]{\includegraphics[width=0.44\textwidth]{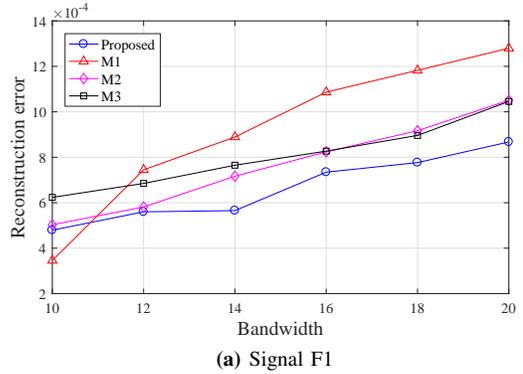}
	\label{error_sensor}}\\
  \centering
	\subfloat[Signal F2]{\includegraphics[width=0.44\textwidth]{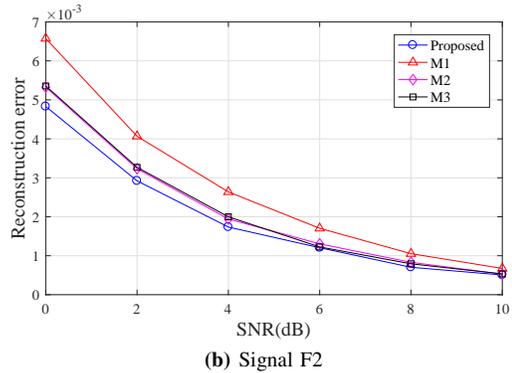}%{snr_sensor_cut}
	\label{snr_sensor}}
\caption{Reconstruction results of G2 for different signals.}
\label{RG}
\end{figure}
We generate 200 signals from each of the two signal models on each of the graphs and set the sample size to 4 times of the bandwidth. The parameter $k$ of M2 that controls how closely the estimate the bandwidth of a signal is set to $8$ in our simulation. 
%\begin{itemize}
%	\item[1)]
	For noisy signal model F1, our method has a better performance in reconstruction error than the others with the same sample size in most case as demonstrated in Fig. (\ref{error_SW}) and Fig. (\ref{error_sensor}). This is because M1 and M2 are heuristic methods, which can not ensure a global optimal solution. M3 formularizes the problem as an experimental design problem like what we do but not allows multiple-time sampling on any node. So their solution is a suboptimal solution from the view of measurement, while we try to find the optimal one. The quantization step is a major factor that influences the performance of our method. The smaller sample size $M$ is, the larger the quantization step is. As a result, the corresponding reconstruction performance will be poorer.
%	\item[2)]
	Our method also has a better performance for signal model F2, especially when the SNR of samples is low as demonstrated in Fig. (\ref{snr_SW}) and Fig. (\ref{snr_sensor}). This also proves that our method is more robust against additive noise.

\section{Conclusion}
In this paper, an algorithm is proposed to obtain the best sampling set for bandlimited GS estimation for noisy samples. By the method of experimental design, we find out the sample quota for each node. Meanwhile, a loose lower bound of sample size is given to ensure the object function in our method is well-defined. 
%Theoretical analysis and experimental results demonstrate that our method has an excellent performance in reconstruction error
% and is more robust against additive noise. 
%Future work would involve considering a local-set-based algorithm for large scale graphs. Another direction would be to extend the signal to the time domain.


\section*{Acknowledgment}
This work is supported by the NSF of China (Grant No. 61501124).




\vfill\pagebreak

%\section{REFERENCES}
%\bibliographystyle{IEEEtran}
%\bibliography{IEEEabrv,references}
%\bibliography{IEEEabrv,references}

\bibliographystyle{IEEEbib}
\bibliography{strings,refs}
%\begin{thebibliography}{1}
%\end{thebibliography}

\end{document}